\documentclass[aps,prb,reprint,superscriptaddress,longbibliography,floatfix]{revtex4-1}

\usepackage{amsmath}
\usepackage{amssymb}
\usepackage{graphicx}
\usepackage{braket}

\usepackage[colorinlistoftodos,prependcaption]{todonotes}

\usepackage{xargs}
\newcommandx{\greencom}[2][1=]
{\todo[inline, color=green!40,#1]{#2}}
\newcommandx{\bluecom}[2][1=]
{\todo[inline, color=blue!40,#1]{#2}}
\newcommandx{\bluemargin}[2][1=]
{\todo[color=blue!40,#1]{#2}}

\begin{document}

\title{Molecular optomechanics in the anharmonic cavity-QED regime using hybrid metal-dielectric cavity modes}

\author{Mohsen Kamandar Dezfouli}
\email{m.kamandar@queensu.ca}
\affiliation{Department of Physics, Engineering Physics and Astronomy,
Queen's University, Kingston, ON K7L 3N6, Canada}

\author{Reuven Gordon}

\affiliation{Department of Electrical and Computer Engineering,
University of Victoria, Victoria, BC V8W 3P2, Canada}

\author{Stephen Hughes}
\email{shughes@queensu.ca}
\affiliation{Department of Physics, Engineering Physics and Astronomy,
Queen's University, Kingston, ON K7L 3N6, Canada}

%%%%%%%%%%%%%%%%%%%%%%%%%%%%%%%%%%%%%%%%%%%%%%%%%%%%%%%%%%%%%%%%%%%%%

%%%%%%%%%%%%%%%%%%%%%%%%%%%%%%%%%%%%%%%%%%%%%%%%%%%%%%%%%%%%%%%%%%%%%

\begin{abstract}
Using carefully designed hybrid metal-dielectric resonators, we study molecular optomechanics in the strong coupling regime
($g_{\rm }^2/\omega_m {>} \kappa$), which 
manifests in anharmonic emission lines in the sideband-resolved region  of the cavity-emitted spectrum ($\kappa{<}\omega_m$).
This nonlinear optomechanical strong coupling regime is enabled through a  metal-dielectric cavity system that yields not only deep sub-wavelength plasmonic confinement, but also dielectric-like confinement times that are more than two orders of magnitude larger than those from typical localized plasmon modes. 
These hybrid metal-dielectric cavity modes enable one to study new avenues of quantum plasmonics for single molecule Raman scattering.
\label{abst}
\end{abstract}
\maketitle

\vspace{0.3cm}

Photons interacting with molecules can induce spontaneous Raman scattering~\cite{Raman}, where optical fields couple to molecular vibrations and scatter at phonon-shifted frequencies with respect to the original excitation frequency. Although most Raman experiments involve very small scattering cross-sections of around $10^{-30}{-}10^{-25}\,{\rm cm^{2}}$, using surface-enhanced Raman spectroscopy (SERS) with metal nanoparticles (MNPs)  \cite{SensingWaveguides,WilletsReview,KneippReview}, enhancement factors of up to $10^{14}$ can be obtained. Carefully fabricated MNPs allow extreme enhancement of electromagnetic fields, in the form of localized hot-spots, which has enabled SERS to emerge as a powerful tool in identifying the structural fingerprint of different molecules and proteins, down to the single molecule level~\cite{Nie1997,Kneipp1997PRL,EtchegoinReview,ChemicalMapping,SeeingSingleMolecule}. MNPs have also been used in hybrid metal-dielectric platforms to optically observe single atomic ions~\cite{Volmer2016}. The SERS process can be viewed as an effective enhancement of the optomechanical coupling between the localized surface plasmon resonance and the vibrational mode of the molecule, which has inspired recent ideas of  molecular optomechanics~\cite{Roelli,Schmidt}. There has also been intense interest in using MNPs to explore regimes of quantum optical plasmonics~\cite{TameQuantumPlasmonics,doi:10.1117/1.JNP.10.033509,doi:10.1021/acs.nanolett.7b02222,StrongCoupling,ControllingSE,LargePF}, including recent work on pulsed molecular optomechanics \cite{PulsedOptomechanics} that observed a superlinear Stokes emission spectrum. However, sufficiently strong optomechanical coupling at the few photon regime, which facilitates nonlinear quantum optical effects such as the single photon blockade~\cite{Rabl,Single-PhotonOptomechanics}, remains largely unexplored in the context of SERS.

A significant problem with MNPs for enhancing quantum light-matter interactions is that considerable metallic losses are involved. Indeed, in stark contrast to dielectric cavity systems, the quality factors for MNPs are only around $Q {\sim} 10$, resulting in
significant cavity decay rates $\kappa$ ($\kappa{=}\omega_c/Q$, with $\omega_c$ the cavity resonance frequency), which is typically much larger than the linewidth of the higher lying quantum state
resonances. This large metallic dissipation can inhibit SERS from probing strong-coupling-like optomechanical resonances, which usually requires a sufficiently sharp spectral change in the optical density of states. Indeed, for most plasmonic resonators, it is convenient to use a modified quantum theory of SERS where the plasmonic system is safely treated as an effective photonic bath~\cite{KamandarACS}. It has also been suggested, using a quasi-static theory, that perhaps there is a universal scaling for the intrinsically low $Q$s of plasmonic resonators \cite{Wang2006}. Thus, it is not surprising that the quantum signatures of molecular optomechanics and SERS under both strong coupling and high quality factors remains relatively unexplored. Such interactions may allow one to study optomechanics in the regime of cavity-quantum electrodynamics (cavity-QED), which requires both the cavity mode and the vibrational mode to be treated quantum mechanically, and without adiabatic elimination. On the other hand, hybrid plasmonic devices, consisting of dielectric and metal parts,  can offer extra design flexibility in terms of the resonance line shapes and cavity mode properties \cite{Barth2010,Eter2014,Doeleman,HybridSERS,hybrid}. Although these hybrid systems involve a more complex coupling than simple MNPs or dielectric cavities, they can be advantageous for quantum plasmonics, as we will demonstrate below. In addition,  cavity mode theory can be reliably used~\cite{hybrid} to extract necessary parameters for cavity-QED studies of these hybrid devices, where the electromagnetic modes can take on the useful properties of both metal and dielectric systems.

\begin{figure}[htb]
\includegraphics[width=0.8\columnwidth]{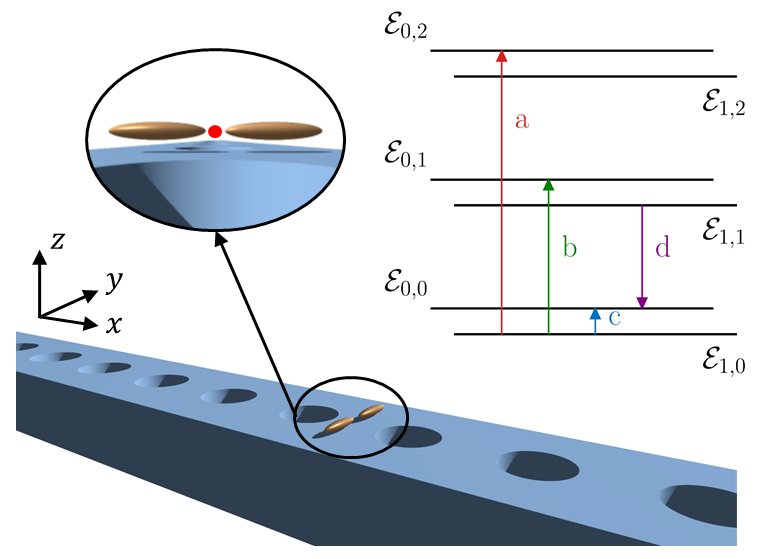}
\caption{Schematic of a hybrid metal-dielectric device with a MNP dimer coupled to a photonic crystal nanobeam cavity, where the inset shows a closeup of the hybrid hot-spot where molecules (shown by a red sphere, though of course they can have a much more general shape) can be trapped. A partial energy diagram of the coupled molecule-cavity system is also shown, where red/green/blue/purple show four transition lines for the cavity emitted  spectrum. 
Note that these quasi-energies are in the interaction picture, where the laser is tuned in resonance with the cavity mode of inetrest, i.e. $\omega_L{=}\omega_c$.
\label{fig:schematic}}
\end{figure}

In this work, we demonstrate how hybrid metal-dielectric systems (schematically shown in Fig.~\ref{fig:schematic}) can probe anharmonic quantum transitions in strongly-coupled molecular optomechanical systems, that are otherwise obscured by the usual metallic dissipation rates. This challenges known limitations of current SERS schemes (which typically probe resonances at the harmonic Stokes and anti-Stokes levels), and opens up possible new avenues in quantum optomechanics using plasmonics. In our hybrid cavity system, there are two dominant hybrid modes that inherit characteristic from both of the dielectric and metallic parent modes, yielding two commonly desired (and/or necessary) properties for our study, namely, sub-wavelength spatial localization (or small effective mode volume, $V_{c}$) and sufficiently small decay rates, $\kappa$ (or sufficiently high $Q$). We stress that both of these two cavity mode features are required to facilitate the strong molecular optomechanics studied in this work.
Specifically, one requires $g_{\rm}^2/\omega_m{>}\kappa$, as well as $\omega_m{>}\kappa$, which is not feasible
with typical plasmonic modes; the latter criterion is a necessary condition to be in the sideband-resolved regime.
For our hybrid system, one of the hybrid modes is plasmonic-like, which maintains the smaller mode volume, while the other mode is dielectric-like, and inherits the larger $Q$ (or smaller $\kappa$). The high $Q$ mode (smaller $\kappa$) of the coupled system can be designed to realize light confinement times of more than two orders of magnitude larger than regular plasmonics resonances ($Q=3500$ compared to $Q\sim10$), and yet maintains a much smaller mode volume well below the diffraction limit in dielectric systems (such as: $V_c{=}5.36\times10^{-6}\,\lambda^3$). 
Such strong mode confinement provided by the hybrid cavity-mode can result in optomechanical coupling rates of $g{=}0.1{-}4\,{\rm meV}$, where the largest value of $g{=}4\,{\rm meV}$ corresponds to $g/\kappa{=}9$, $g^2/\omega_m\,\kappa{\approx}3.5$ and $g/\omega_m{=}0.4$ for the common figures of merit for strong and ultrastrong optomechanical coupling. We subsequently use the high $Q$ hybrid mode (small $\kappa$) of our hybrid device to demonstrate the regime of strong optomechanical coupling, in which pronounced shifts of the cavity frequency as well as new anharmonic Raman side-peaks (first and higher-order Stokes and anti-Stokes resonances) are observable in the cavity emitted spectrum. The low $Q$ mode is also interesting in its own right, but for this work we explore that the high $Q$ mode to access the sideband-resolved regime.

The layout of the rest of our paper is as follows. First, we describe the optomechanical Hamiltonian without any form of linearization and include a coherent cavity pump term. In the weak pumping regime, we discuss the analytical structure
of the optomechanical states and eigenenergies, which are well known. We additionally present quantum master equations that can be used to obtain the system dynamics and emitted spectrum under the influence of system-bath dissipation channels. This includes both standard and generalized master equations~\cite{ModifiedME}.
Next, we present a classical modal analysis (using QNMs) of metallic dimers of different gap sizes as well as the hybrid metal-dielectric system, and extract the necessary cavity parameters used in this study such as effective mode volume and quality factor. Then, we present our quantum simulation results including the coupled-cavity emitted spectrum and show the role of temperature as well as the additional dissipation beyond the standard master equation, on the emitted spectrum. We also show how laser detuning qualitatively affects the mode populations and emission spectra. Finally, we present our conclusions and closing remarks.

\section*{Molecular optomechanics under strong coupling}\label{sec:hamiltonian}

\subsection{Optomechanical system Hamiltonian, eigenenergies and dressed states}

Below we wish to probe strong coupling effects beyond the usual
coupled oscillator model. For example, with regards to probing vibrational strong coupling effects
with molecules, del Pino {\it et al.}~\cite{delPino2015}
have studied collective strong coupling between vibrational excitations and  an infrared cavity modes, using the  
standard Rabi interaction Hamiltonian
$\hbar g_{\rm Rabi}(a+a^\dagger)(b+b^\dagger)$,
and they
also considered resonant Raman interactions;
here $a,a^{\dagger}$ and $b,b^{\dagger}$ represent the 
creation, annihilation operators of the cavity and vibrational modes, respectively.
For $g_{\rm }{>}\kappa$, where $\kappa$ is the cavity decay rate, they found strong coupling features
in the cavity emitted spectrum, and also studied effects
associated with the ultrastrong coupling regime (USC), namely when $g_{\rm }/\omega_m{>}0.1$. Indeed, they found that the $n$th Stokes line splits into $n + 1$ sidebands. 
In the regime of molecular optomechanics,
the same form of interaction Hamiltonian
can be realized  through a linearization procedure
~\cite{RevModPhys.86.1391,Roelli,Schmidt,PulsedOptomechanics,KamandarACS},
so that $g  a^\dagger a (b+b^\dagger) \rightarrow
g' (a+ a^\dagger) (b+b^\dagger)$,
where $g'= \alpha g$, and
$\alpha$ is the displaced amplitude~\cite{Schmidt}.

For our study, we employ the  fundamental optomechanical interaction without any form of linearization, including the cavity pump field in the interaction picture:~\cite{RevModPhys.86.1391}
\begin{align}
H_{\rm s} & =\hbar\Delta\,a^{\dagger}a+\hbar\omega_{m}\,b^{\dagger}b-\hbar g\,a^{\dagger}a\left(b^{\dagger}+b\right)+\hbar\Omega\left(a^{\dagger}+a\right),\label{eq:H}
\end{align}
where $\Delta=\omega_{c}-\omega_{L}$ is the detuning between the
optical cavity and the pump laser, $\omega_{m}$ is frequency of molecular vibrational mode, and $\Omega$ is the Rabi frequency of the optical cavity mode (and we have made a rotating wave approximation for terms rotating at $e^{2i\omega_L t}$). Note that cavity operator terms $aa$ and $a^{\dagger}a^{\dagger}$ can be safely ignored
here, as $\omega_c\gg \omega_m$\cite{RWA1,RWA2}.
The optomechanical coupling factor is  $g=\left(\hbar R_{m}/2\omega_{m}\right)^{-1/2}\omega_{c}/\varepsilon_{0}V_{c}$ \cite{Roelli}, with $R_{m}$  the Raman activity associated with the vibration under study, and $V_{c}$ as the effective mode volume of the cavity mode under investigation.

The optomechanical coupling term in Eq.~\eqref{eq:H} is appropriate for describing off-resonant Raman interactions. For resonant interactions, the plasmonic MNP also interacts with electronic (two-level) vibrational degrees of freedom, through~\cite{delPino2015}
$\omega_m \sqrt{S}\sigma^+\sigma^-(b^\dagger+b)$\footnote{Note that this work actually considers the resonant interaction between a cavity mode and collective vibrational modes, but they also include resonant Raman interactions terms.},
where $\sigma^+,\sigma^-$ are the Pauli operators and
$S$ is the Huang-Rhys parameter, which quantifies the phonon displacement between the ground and excited electronic states.
As discussed in Ref.~\onlinecite{SERS}, resonant Raman effects may be treated phenomenologically, resulting 
in an effective increase of the off-resonant interaction above. However, this is likely only a good approximation for weak pumping fields, where the Fermionic operators behave as harmonic oscillator states. For our studies below, we concentrate on the off-resonant Raman interactions but also use Raman cross sections that can likely be boosted using resonant Raman interactions.

Neglecting the influence of the cavity pump term for now, analytical insight into the resonances of the SERS Hamiltonian of Eq.~\eqref{eq:H}, can be obtained. In the interaction picture, the system dressed-state energies take the 
form~\cite{agarwal_quantum_2012}:
\begin{equation}
{\cal E}_{n,k}= n\hbar\Delta+ k\hbar\omega_m
-n^2 \frac{\hbar g^2}{\omega_m},
\label{eq2}
\end{equation}
with the corresponding
eigenstates,
\begin{equation}
\ket{\Psi_{n,k}}=
{D}^\dagger\left (\frac{gn}{\omega_m}\right)
\ket{n,k},
\label{eq3}
\end{equation}
where ${D}$ is the displacement operator. Thus the optomechnical eigenstates in the phonon space are obtained by displacing number states $\ket{k}$ for phonons, which in turn also depends on the number of photons (through $n$).
Specifically, the phonon states are displaced as follows: $b\rightarrow b{-}d_0 a^\dagger a/\omega_m$, where $d_0{=}g/\omega_m$ is the normalized displacement. This results in photon manifolds that contain phonon sub-levels, where the sub-level splitting depends on the photon number state. For example, 
within the $n{=}0$ photon manifold, one can probe the usual Raman sidebands,
$\pm \omega_m, \pm 2 \omega_m, etc$;
from $n{=}1$ to $n{=}0$, then one can explore Stokes resonances
at  $\omega_c{\pm}- g^2/\omega_m
-k\omega_m$, and even
richer anharmonicities from the
higher photon manifolds (which will be harder to resolve).
To resolve the lowest-order anharmonic levels ($n{=}1$ photon manifold),  one needs to meet the condition
$g^2/\omega_m {>} \kappa$, which gives us one of the criteria for optomechanical strong coupling. On the other hand, one also requires
$\kappa {<}\omega_m$, to be able to
be in the sideband resolved regime~\cite{Neumeier2018}.
We also stress that such optomechanical states also involve interactions
well beyond the rotating-wave approximation for the mechanical mode, and usually one also finds that
$g/\omega_m{>}0.1$, which is characteristic of the
USC regime~\cite{USCR1,Niemczyk2010,ModifiedME,delPino2015}.
Note the same displacement occurs using a
polaron transform of the system Hamiltonian,
which has an identical polaron shift appearing
for the first excited state photon manifold~\cite{MahanBook,Rabl,NeumanThesis}.
However, for the cavity photon operators (bosons),
it also introduces a nonlinear Kerrlike
term ($ \propto -g^2/\omega_ma^\dagger a^\dagger a a$), which can be difficult to
account for numerically.

Figure~\ref{fig:schematic} shows, schematically, four of the optomechanical energy levels which can be resolved in the emitted spectrum that we discuss later. For $\Delta{=}0$, the first three energy levels for $n{=}0$ are ${\cal E}_{0,0}{=}0$, ${\cal E}_{0,1}{=}\hbar\omega_m$ and ${\cal E}_{0,2}{=}2\hbar\omega_m$ for the ground state, first order and second order 
vibrational states. With an applied field, emission at phonon
peaks of $m\omega_m$ on the red and blue side of the cavity resonance cause
 the standard Stokes and anti-Stokes emissions. 
 However, for the optomechnical dresses states,
 the first three energy levels for the $n{=}1$ photon manifold
contain the anharmonic side-bands, i.e., ${\cal E}_{1,0}{=}{-}\hbar g^2/\omega_m$, ${\cal E}_{1,1}{=}\hbar\omega_m{-}\hbar g^2/\omega_m$ and ${\cal E}_{1,2}={2}\hbar\omega_m{-}\hbar g^2/\omega_m$. Note that these are all shifted by the same amount with respect to standard Raman emissions. They also involve changes for both cavity and molecule, and therefore represent the optomechanical feedback between the two coupled oscillators. While the ${-}\hbar g^{2}/\omega_{m}$ spectral shift depends on the strength $g$, it does not explicitly depend on the cavity quality factor, $\kappa$ (using the simple analysis above). However, one also requires $\kappa{<}\omega_m$ when dissipation is included. This is important, because the plasmonic-like modes have $\kappa{>}\omega_m$, and would fail to resolve such states in general. 

\subsection{Quantum master equations}

With cavity and mechanical (vibrational) bath interactions included, we first employ a standard master equation approach \cite{Breuer,Carmichael}, which is used  to compute different observables of interest such as population dynamics and the cavity emission spectrum. The ensuing master equation is \cite{Schmidt,KamandarACS}
\begin{align}
\frac{d\rho(t)}{dt} & =-\frac{i}{\hbar}\left[H_{\rm s},\rho(t)\right] + \frac{\kappa}{2}\,\mathcal{D}[a]\rho(t)\nonumber\\
 & +\frac{\gamma_{m}\left(\bar{n}^{{\rm th}}+1\right)}{2}\,\mathcal{D}[b]\rho(t) + \frac{\gamma_{m}\bar{n}^{{\rm th}}}{2}\,\mathcal{D}[b^{\dagger}]\rho(t),
\label{eq:rho}
\end{align}
where $\kappa$ is the cavity decay rate, $\gamma_{m}$ is the vibrational decay rate, $\bar{n}^{{\rm th}}$=$\left({\rm exp}\left(\hbar\omega_{m}/k_{B}T\right)-1\right)^{-1}$ is the thermal population of the vibrational mode at temperature $T$, with $k_{B}$  the Boltzmann constant, and the Lindblad superoperator $\mathcal{D}$ is defined via:
$\mathcal{D}[O]\rho(t)=2O\rho(t) O^{\dagger}-O^{\dagger}O\rho(t)-\rho(t) O^{\dagger}O$.

One potential problem with the standard master equation
is that it neglects internal coupling
between the system operators when deriving
the system-bath interactions.
This general problem was discussed
in 1973 by Carmichael and Walls~\cite{Carmichael1973}, where they showed that
the correct bath interaction should occur
at the dressed resonances of the system. This
``internal coupling'' interaction has been applied to
a wide variety of problems, including 
Mollow triplets with plasmon resonators  interacting with two level atoms~\cite{Ge2013},
circuit QED~\cite{Beaudoin2011}, and general regimes
of USC physics~\cite{RWA2}. 
Excluding the weak pumping field, the
dynamics of $b$ and $a$ can be solved analytically
from the system Hamiltonian, which allows one to obtain a self-consistent solution for the dissipation terms.
Neglecting terms that
oscillate at $\exp(\pm i\omega_m t)$
and $\exp(\pm i2\omega_m t)$ (in the interaction picture), the solution
has been derived by Hu {\em et al.}~\cite{ModifiedME}, and takes the form
\begin{align}
\frac{d\rho(t)}{dt} & =-\frac{i}{\hbar}\left[H_{\rm s},\rho(t)\right] + \frac{\kappa}{2}\,\mathcal{D}[a]\rho(t)\nonumber\\
 & +\frac{\gamma_{m}\left(\bar{n}^{{\rm th}}{+}1\right)}{2}\,\mathcal{D}[b{-}d_0 a^{\dagger}a]\rho(t) \nonumber\\
 & + \frac{\gamma_{m}\bar{n}^{{\rm th}}}{2}\,\mathcal{D}[b^{\dagger}{-}d_0a^{\dagger}a]\rho(t)  + \frac{2\gamma_{m}k_{\rm B}{\rm T}d_0^2}{\hbar\omega_m}\mathcal{D}[a^{\dagger}a]\rho(t),
\label{eq:rho2}
\end{align}
which has no affect on the final cavity decay terms, but causes the mechanical dissipation to be displaced. There is also an additional pure dephasing process 
which we have numerically checked to be negligible in the regimes below (though we include it above for completeness).
The origin of the  dissipation modifications stem  from
$b(t){=} e^{i H_{\rm s}/\hbar\, t}\, b\, e^{-i H_{\rm s}/\hbar\, t} {=} (b-d_0 a^\dagger a)e^{i\omega_mt} + d_0 a^\dagger a$. In a standard master equation,
one usually assumes $b(t){\approx}b$ when deriving
the system-bath interactions terms, which typically fails
in the USC regime or in regimes that probe dressed states that are sufficiently far from the laser frequency.
To demonstrate the additional physics behind this modified dissipation,  below we start by using the commonly used master equation of Eq.\,\eqref{eq:rho} (standard master equation), and then also carry out a direct comparison with Eq.\,\eqref{eq:rho2} (corrected, or dressed-state, master equation) to show any modifications that are introduced from the more correct dissipation terms.

Since we are dealing with lossy mode systems,
where the loss is substantial,
it is also worth noting that while there has been some recent progress made with quantizing  QNMs for any open system~\cite{franke_quantization_2018}, we will neglect the additional complexities for this work, and choose designs where this is expected to be a  good approximation (namely, in the regime of a single mode master equation~\cite{franke_quantization_2018}).

For numerical calculations
of the  quantum master equation, and for calculating the cavity emitted spectrum, we employ the qutip package~\cite{qutip,qutip2}, under Python. We performed a basis analysis in terms of Hilbert space size, and confirmed that including up to $\left|n{=}6,k{=}6\right\rangle$ states leads to numerically converged spectrum, for the full range of parameters considered below.

\section*{Electromagnetic modal analysis and molecular optomechanical parameters}\label{sec:qnm}

We first employ the full
three-dimensional  Maxwell equations to  design and understand a suitable hybrid metal-dielectric system. Our goal is to design a cavity mode with a very small (plasmon-like) mode volume and a suitably small dissipation rate (large $Q$).
Specifically, we consider a MNP dimer that is top-coupled to a photonic crystal nanobeam cavity (see Fig.~\ref{fig:schematic}). Similar systems have been discussed before \cite{hybrid} and made~\cite{Mukherjee2011}, but here we significantly improve the design for this study, namely for anharmonic strong coupling, which requires a larger $Q$ and much smaller $V_{ c}$.
We do this by adjusting the gap size, the aspect ratio, and the shape of the plasmonic dimer as well as its spacing from the dielectric nanobeam. The latter in practice can be implemented using spacer layers, where the refractive index change causes resonance frequency shifts for the main plasmonic resonance. This can be easily compensated for by small adjustments on the dimer aspect ratio, however, we find it of less influence on the high $Q$ mode of the hybrid design that we are interested in and so neglect the details of adding spacer layers below. Interesting alternative metal-dielectric hybrid structures have been discussed by Doeleman {\em et al.}~\cite{Doeleman}.

\begin{figure*}[htb]
\includegraphics[trim={0.2cm 0.2cm 1.5cm 1.5cm},clip,width=\textwidth]{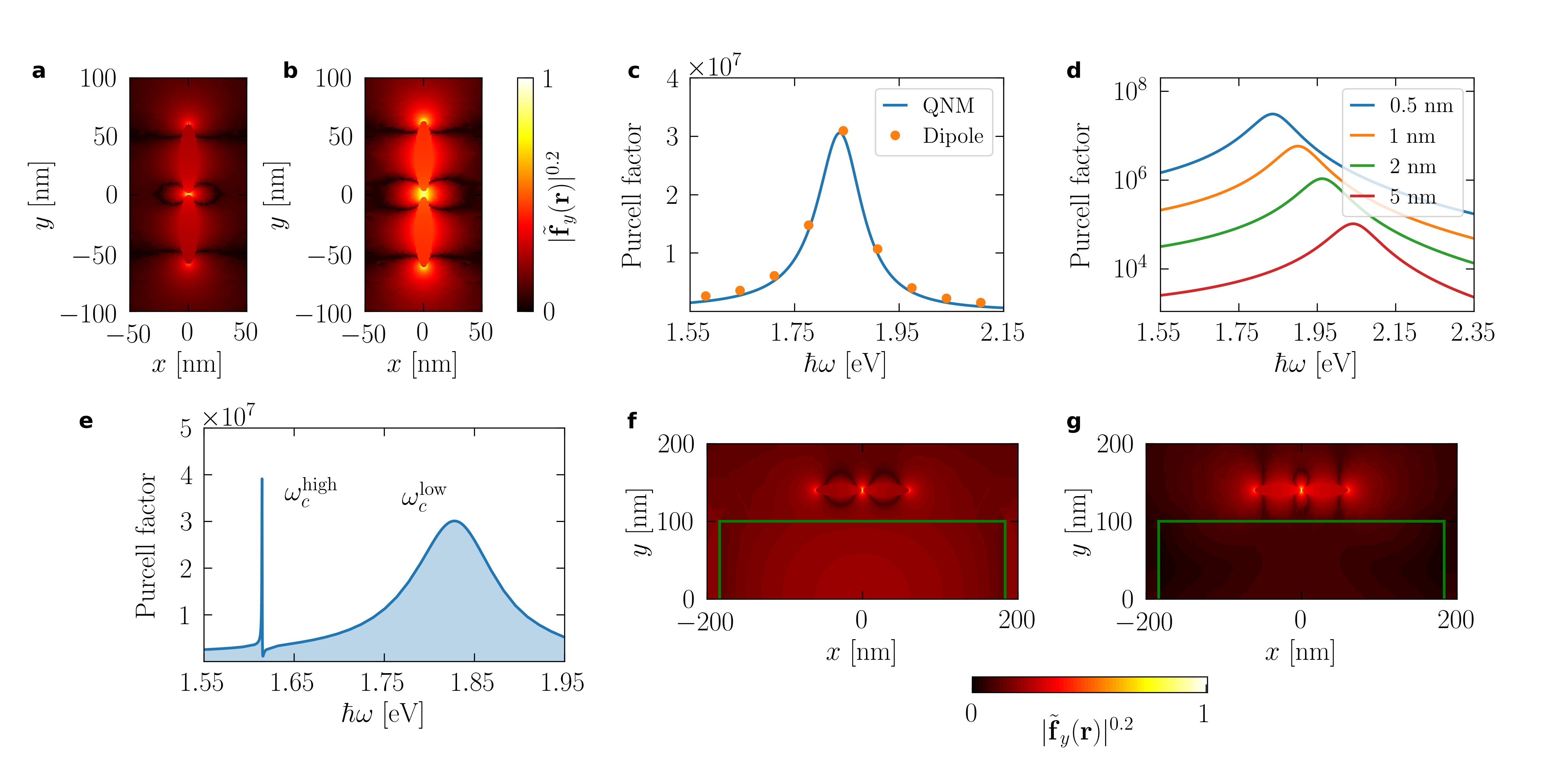}
\caption{{\bf a}-{\bf b} Calculated mode profile of two gold dimers with different gaps,  0.5 nm and 5 nm, respectively. {\bf c} Purcell factor
calculations for the smaller gaps size, 0.5 nm, using the analytical QNM theory (solid curve)
and fully vectorial solutions to Maxwell's equations from a dipole excitation (symbols). {\bf d} QNM calculated Purcell factor for dimer designs with different gap sizes. {\bf e} Purcell factor of a dipole emitter placed inside the plasmonic gap of the hybrid device, where a sharp high $Q$ mode (dielectric-like) is present next to the broader low $Q$ mode (plasmonic-like). {\bf f}-{\bf g} Calculated QNM profile ($yz$-cut) for the high $Q$ and low $Q$ modes of the hybrid system, where the green rectangle shows the boundary of the dielectric beam. Note that, for all the mode profiles, a nonlinear color scaling ($|{\bf f}_y|\rightarrow|{\bf f}_y|^{0.2}$) is used for better visualization.\label{fig:pf}}
\end{figure*}

For our specific MNP design,
we consider a gold plasmonic dimer that is made of two ellipsoids, each $60\,{\rm nm}$ long and $15\,{\rm nm}$ wide. A small gap, ranging from 0.5-5\,nm in between them is used to create a pronounced field hot-spot for trapping molecules. Notably, the smallest gap used here, though likely very challenging, has been experimentally achieved \cite{05nmGap} and it stays within the region where electron tunneling effects are negligible \cite{QuantumPlasmonicsReview}. Additionally, using a
fully three-dimensional nonlocal QNM theory at the level of a hydrodynamical model~\cite{KamandarDezfouliOptica}, we have confirmed that the nonlocal considerations mostly just blueshift the low $Q$ mode of our hybrid device by about 2\%, which again, can be tuned back by adjusting the dimer aspect ratio. The MNP dielectric function is thus modeled using a local Drude theory, $\varepsilon_{{\rm MNP}}\left(\omega\right)$=$1-\omega_{p}^{2}/\omega\left(\omega+i\gamma_{p}\right),$ with plasmon energy of $\hbar\omega_{p}$=$8.2934\,{\rm eV}$ and collision broadening of $\hbar\gamma_{p}$=$0.0928\,{\rm eV}$.

The classical mode calculations are performed using the commercial frequency-domain solver from COMSOL, and the QNMs are calculated using the technique presented in Ref.~\onlinecite{Bai2013}.
The QNMs are poles of the electic-field photonic Green function and they can be used to accurately estimate the effective mode volumes at the emitter location as well as the quality factors. 
The QNM complex eigenfrequencies are
defined from~\cite{PhilipACS} $\tilde \omega_c {=} \omega_c{-}i\gamma_c$,
where $\kappa{=}2\gamma_c$, $Q{=}\omega_c/\kappa$ and the effective mode volume is obtained from the normalized QNM spatial profile
at dimer gap center~\cite{PhilipACS}, $V_c={\rm Re}\{1/\epsilon_{\rm B} \tilde{\bf f}^2({\bf r}_0)\}$,
where $\epsilon_{\rm B}$ is the background dielectric constant in which the molecule is located (assumed to be 1 here). 

For our detailed calculations, 
we model both metal dimer structures on their own and dimers on top of
photonic crystal cavity beams (see Fig.~\ref{fig:schematic}).
For the dimer calculations, a computational domain of $1\,{\mu m}^3$ was used with 10 layers of perfectly matched layers and a maximum mesh size of 3 nm over the metallic region. All our mode calculations are fully three dimensional.
For the hybrid coupled-cavity system, a computational domain of $100\,{\mu m}^3$ was used with the additional requirement of the maximum mesh size of 50 nm over the dielectric beam.
The nanobeam cavity is assumed to be made of silicon-nitride with a refractive index of $n$=$2.04$, with height ${h}$=$200\,{\rm nm}$, and  width ${w}$=$367\,{\rm nm}$. A dielectric cavity region that is $126\,{\rm nm}$ long is created in the middle of the nanobeam, with first a taper section and then a mirror section on either side. The taper section is made of $7$ holes linearly increased from $68\,{\rm nm}$ to $86\,{\rm nm}$ in radius, and from $264\,{\rm nm}$ to $299\,{\rm nm}$ in spacing. The mirror section, however, is made of $7$ more holes having the same radius of $r{=}86\,{\rm nm}$ and the same spacing of  $a{=}306\,{\rm nm}$. For the main mode of interest for this cavity design, we obtained the effective mode volume at the beam center inside dielectric region to be $V_{c}{=} 0.078\,\lambda^3$, with the corresponding quality factor of $Q{=}3 {\times} 10^{5}$.

Figure \ref{fig:pf} summarizes how the dimer mode properties change as a function of gap size (a-d), and also shows the hybrid mode properties for the smallest gap sizes on a photonic crystal beam (e-g).
The results in Figs.~\ref{fig:pf}(a-b) show 
a surface plot of the near-field mode profile for gap sizes of 0.5~nm and 5~nm, respectively. 
In Fig.~\ref{fig:pf}(c), we show the reliability of our implemented QNM theory in accurately capturing the system response by comparing against full dipole solutions of Maxwell's equation (i.e., with no approximations). The dipole is polarized
along the dimer axis, namely along $y$. Remarkably, these
results  show that even for gap sizes as small as 0.5\,nm, a single QNM gives an excellent fit compared to full dipole simulations for the enhanced emission factor.
Figure \ref{fig:pf}(d) shows how the QNM calculated Purcell factor~\cite{PhilipACS} (enhanced emission rate from a dipole) 
changes for several dimer designs with different gap sizes, ranging from $0.5\,{\rm nm}$ to $5\,{\rm nm}$, while every other geometrical parameters aside from the gap size are kept the same. As seen, decreasing the dimer gap induces a resonance redshift as well as a dramatic increase in the local density of states, which translates into a decreasing effective mode volume.

\begin{table}[htb]
\begin{centering}
\begin{tabular}{|c|c|c|c|}
\hline 
 device/mode & $V_{ c}/\lambda_c^3$ & $\hbar\kappa$ [meV]  & $\hbar g$ [meV] \tabularnewline
\hline 
\hline 
0.5 nm gap dimer & $4.38\times10^{-8}$ & 105 & 24.96 \tabularnewline
\hline
1 nm gap dimer & $2.35\times10^{-7}$ & 107 & 5.31 \tabularnewline
\hline
2 nm gap dimer & $1.29\times10^{-6}$ & 108 & 1.10 \tabularnewline
\hline
5 nm gap dimer & $1.36\times10^{-5}$ & 111 & 0.12 \tabularnewline
\hline
high $Q$ hybrid & $5.36\times10^{-6}$ & 0.46 & 0.10 \tabularnewline
\hline
low $Q$ hybrid & $4.53\times10^{-8}$ & 108 & 23.27 \tabularnewline
\hline
\end{tabular}
\par\end{centering}
\caption{Calculated mode volume, decay rate and the optomechanical coupling factor for various dimer designs, as well as the two dominant modes of the studied hybrid device. Coupling values are calculated for a low frequency oscillation at $\hbar\omega_{m}$=$10\,{\rm meV}$ holding a Raman activity of $R_{m}$=$10^{3}\,{\rm \mathring{A}^{4}{\rm amu}^{-1}}$, as discussed in the main text.\label{tab:veff}}
\end{table}

Placing the dimer on top of the nanobeam cavity forms a platform that can trap molecules inside the hot-spot gap, in the presence of dielectric mode couping (see Fig.~\ref{fig:schematic}). In Fig.~\ref{fig:pf}(e), we show the total Purcell factor for a dipole emitter that is again oriented along the dimer axis with photonic crystal cavity coupling. As seen, over the wide range of $400\,{\rm meV}$, only two modes contribute dominantly. These are the two hybrid cavity modes also calculated rigorously using QNM theory~\cite{hybrid,Bai2013,PhilipACS}. 
The resonant frequencies of the hybrid modes are found to be $\hbar\omega_{c}^{{\rm HQ}}$=$1.61\,{\rm eV}$
and $\hbar\omega_{c}^{{\rm LQ}}$=$1.83\,{\rm eV}$, with corresponding quality factors of $Q^{{\rm HQ}}$=$3500$ and $Q^{{\rm LQ}}$=$17$, respectively. Additionally, the effective mode volumes for the two hybrid modes are $V_{c}^{{\rm HQ}}$=$5.36\times10^{-6}\,\lambda_{c}^3$ and $V_{c}^{{\rm LQ}}$=$4.54\times10^{-8}\,\lambda_{c}^3$. Note that, as  discussed before, the high $Q$ mode has an extremely small mode volume inherited from the plasmonic dimer structure. In Fig.~\ref{fig:pf}(f) (g), we show the $yz$ cut of the high (low) $Q$ hybrid QNM profiles, for completeness.

The results of our classical modal investigation are summarized in Table~\ref{tab:veff}, where a relatively constant quality factor is maintained across all devices (apart from the high $Q$ hybrid mode), even though the mode volume changes by several orders of magnitude.
Note that the plasmonic mode has a 
$\sim$200-fold increase in the coupling rate, even though the mode volume is
decreased by $\sim$100 in normalized units. This is caused also by the change in cavity wavelength, which appears
cubed.

For the molecular vibrational mode, we consider a reasonably low frequency oscillation at $\hbar\omega_{m}$=$10\,{\rm meV}$ that can have a Raman activity of $R_{m}$=$10^{3}\,{\rm \mathring{A}^{4}{\rm amu}^{-1}}$ (these numbers are within the range available in the literature, e.g., for single-walled carbon nanotubes \cite{SWCNT,Roelli}). Indeed, low frequency vibrations enforce the interaction to take place within the close vicinity of the high $Q$ mode and they are expected to have higher Raman cross-sections \cite{lowfreqs}, and therefore can offer higher optomechanical coupling rates (see definition of $g$). While this work was for small dielectric particles, the link between continuum models and molecules levels has been  demonstrated for the fullerine family~\cite{Adhikari2011}, and much lower wavelength  modes have been explored for molecules using  a low-wavenumber-extended confocal Raman microscope~\cite{Lebedkin2011}.

We also consider an intrinsic mechanical quality factor of $Q_m$=$\omega_{m}/\gamma_{m}$=$100$ for the molecular vibration. However, note that the temperature dependence will increase the vibrational linewidth, as incorporated through $\bar n^{{\rm th}}$ in  Eq.~\eqref{eq:rho}. For such a vibration mode, by using the expression for the optomechanical coupling provided earlier, the two hybrid modes yield $\hbar g^{{\rm HQ}}$=$0.1\,{\rm meV}$ and $\hbar g^{{\rm LQ}}$=$20\,{\rm meV}$. As discussed earlier, even though the low $Q$ mode offers a much larger coupling factor, it fails to exploit the strong coupling interaction, since one requires $\kappa{<}\omega_m$. These estimates are made under the standard  off-resonant SERS regime, whereas the resonant Raman cross-sections (on the order of $10^{-25}\,{\rm cm^{2}}$) are typically known to be $10^{5}$ larger compared to standard Raman cross-sections (on the order of $10^{-30}\,{\rm cm^{2}}$). Therefore, the effective coupling factor can likely be enhanced by more than $\times10^{2}$ when resonant Raman regimes are used. 
Thus, it is not too unreasonable for us to increase the
coupling factor to better explain the underlying physics
of an increased interaction rate.
Consequently, we  will consider several values of the coupling factor within the  range $\hbar g^{{\rm HQ}}{=}0.1$-$4\,{\rm meV}$ (where new spectral features are visible at around $2\,{\rm meV}$) to identify new spectral features for molecules coupled to the high $Q$ mode. 
As discussed earlier, the resonant Raman interaction
is fundamentally different to off-resonant Raman, but this approach is expected
to be  reasonable for weak pump fields, where
one can treat the resonant Raman process phenomenologically~\cite{SERS}.
A more detailed investigation of resonant Raman interactions~\cite{delPino2015,NeumanThesis}, increased pump fields,
and vibrational electronic coupling is left to future work.

\section*{Molecular optomechanical spectrum and population dynamics under strong coupling}\label{sec:results}

Employing the quantum master equations (Eq.~\eqref{eq:rho} and Eq.~\eqref{eq:rho2}), and the quantum regression theorem~\cite{Carmichael}, we calculate the cavity emitted spectrum of the hybrid device from a Fourier transform of the first-order quantum correlation functions:
\begin{align}
S\left(\omega\right) & \equiv {\rm Re}\left\{ \int_{0}^{\infty}dt\,e^{i\left(\omega_{L}-\omega\right)t}\right.\label{eq:S}\\
 & \qquad\left.\vphantom{\int_{0}^{\infty}}\times\left[\left\langle a^{\dagger}\left(t\right)a\left(0\right)\right\rangle _{{\rm ss}}-\left\langle a^{\dagger}\right\rangle _{{\rm ss}}\left\langle a\vphantom{a^{\dagger}}\right\rangle _{{\rm ss}}\right]\right\} ,\nonumber 
\end{align}
where the expectation values are taken over the system steady state and the coherent contribution is subtracted off. Note that this emission spectrum is the {\it incoherent} spectrum, stemming from quantum fluctuations about steady-state; this emission spectrum could be detected background free in the same way that quantum dot Mollow triplets are detected when coupled to cavity modes~\cite{Ulhaq2013}, and the coherent contribution from a laser source is much smaller than the spectral features we are probing. We also calculate the population dynamics of the cavity mode from $n_{c}=\braket{a^\dagger a}(t)$, where at time $t{=}0$, the cavity is in the ground state (not populated) and the vibrational mode is in a thermal state, according to the chosen temperature.

\begin{figure*}[htb]
\includegraphics[trim={0.2cm 0.2cm 1.2cm 0.5cm},clip,width=1\textwidth]{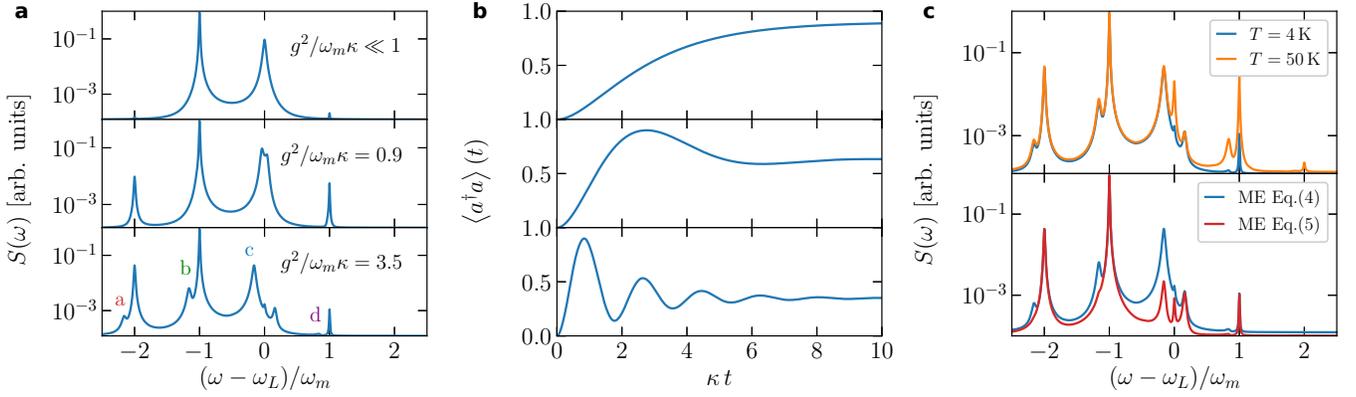}
\caption{{\bf a} Cavity emitted spectrum and {\bf b} cavity photon population versus time, plotted for $\hbar g=0.1,2,4$ meV, top to bottom, respectively. 
The temperature is $T=4\,$K, and for the larger
values of $g$, $g/\omega_m=0.4$. this is also in the USC regime.
The  first (lowest) $g$ value is for the hybrid high $Q$ design shown earlier, with the estimated non-resonant Raman configuration. Here, the cavity decay rate is $\hbar\kappa=0.46\,{\rm meV}$, the frequency of vibration is $\hbar\omega_m=10\,{\rm meV}$, and the Rabi energy is $\hbar\Omega$=$0.1\,{\rm meV}$. Note that $\omega_m \gg \kappa$ in this case, as is required for sideband resolution. {\bf c} Influence of the temperature (top, 4~K versus 50~K) and the corrected Lindbald dissipation terms (bottom) on the strong coupling spectrum for $\hbar g=4\,{\rm meV}$ case at $T=4\,{\rm K}$ (a, bottom).
\label{fig:gscan} }
\end{figure*}

In Figs.~\ref{fig:gscan}(a-b), we show the
cavity-emitted spectrum  as well as the corresponding population dynamics to steady state for the optomechanical coupling rates $\hbar g^{\rm HQ}$=$0.1,\,2,\,4\,{\rm meV}$.
The lowest value of $g$ is our starting point, which is estimated for the off-resonant Raman excitation using the high $Q$ mode. The highest value of $g$ is roughly an order of magnitude smaller than the upper estimate for the resonant Raman, which we think is reasonable given our approximate model.
The simulations in Figs.~\ref{fig:gscan}(a-b) use the standard master equation (Eq.~\eqref{eq:rho})
and assume a  temperature of $T{=}4\,{\rm K}$.
By increasing $g$, we  see 
significant non-trivial shifts of the cavity resonance and the emergence of the associated Raman side-peaks (Fig.~\ref{fig:gscan}(a)). In Fig.~\ref{fig:gscan}(b),  the corresponding cavity populations become non-trivial with respect to time as one gets into the  strong coupling regime.

Based on the earlier analytical discussion of the eigenenergies, it is easy to see where the additional peaks in the full system spectrum originate from, i.e., what kind of transitions they correspond to. Referring to the schematic energy diagram of Fig.~\ref{fig:schematic}, the sidebands are mediated from the anharmonic energy levels introduced by the sufficiently strong optomechanical coupling and involve jumps on the molecular ladder as well as the cavity ladder, showing signatures of the strong coupling. We stress that anything beyond the
first-order Stokes and anti-Stokes resonances, is already 
well into a nonlinear regime that is beyond the usual 
linearization procedure for  optomechanical interactions~\cite{Schmidt,KamandarACS}; furthermore,
the resonances that are not at multiples
of the phonon energy are related to the nonlinear anharmonic cavity-QED regime." Also note the analytical energy eigenstates include photons and phonons to all orders (accessible through the pumping field), though these are significantly modified in the presence of dissipation.

We next explore the effect of increasing temperature on the emission spectrum, in Fig.~\ref{fig:gscan}(c) (top, for  $T$=$4,\,50\,{\rm K}$). The emission spectrum shows that increasing the temperature mainly affects the anti-Stokes emissions. Even at room temperature, the thermal phonon populations for the vibration mode energy of  $\hbar\omega_{m}$=$10\,{\rm meV}$ is about $n^{{\rm th}}$ $\approx 2$ and therefore, a further increase of the anti-Stokes emissions as well as some broadening can occur. 

Note that some of the peaks appearing in the spectrum
are not immediately explained from the previous (pump-free)
eigenvalues and eigenstates in Eqs.~\eqref{eq2}-\eqref{eq3}. For example,
there is clearly a central peak at $\omega_L$ (even for small values of $g$) as well as at $\omega_L{+}g^2/\omega_m$ (see Fig.~\ref{fig:gscan}(c), bottom), and it is tempting to ask if these are associated with higher lying photon states.
In fact, these transitions can be fully explained by including a weak pumping field in the system Hamiltonian, and truncating the Hilbert space to only include one photon and one phonon state (so up to two quanta), which is still beyond weak excitation. In this case, the lowest three ``Floquet eigenenergies''~\cite{Illes:15}
are also obtained analytically (with $\omega_L{=}\omega_c$, and $\Omega{\ll} g$):
${\cal E}^{\rm Fl}_0{=}{-}\hbar g^2/\omega_m$, ${\cal E}^{\rm Fl}_1{=}0$,
${\cal E}^{\rm Fl}_2{=}\hbar\omega_m$. These are quasienergies,
and transitions between the Floquet states cause resonances at
$\omega_L,\omega_L{\pm} \omega_m,\omega_L{\pm}  g^2/\omega_m$
and $\omega_L {\pm} (\omega_m {+}  g^2/\omega_m)$, which explain the additional peaks on the spectrum. Note that the central peak is similar to the central peak of a Mollow spectrum (for a driven two level system), which is also a nonlinear effect from the drive.

Additionally, in Fig.~\ref{fig:gscan}(c) (lower), we study the effect of the modified dissipation terms given in Eq.~\eqref{eq:rho2} (generalized master equation). As seen, under the exact same configuration, the additional Lindblad terms  of the generalized master equation introduce additional dissipation that obscures some of the new side-peaks, consistent with the results of 
Ref.~\onlinecite{ModifiedME}. Furthermore, this extra broadening seems to affect the Stokes emissions more than the anti-Stokes emissions.

\begin{figure}[ht!]
\includegraphics[trim={0.cm 0.2cm 0.5cm 0.5cm},width=.6\columnwidth]{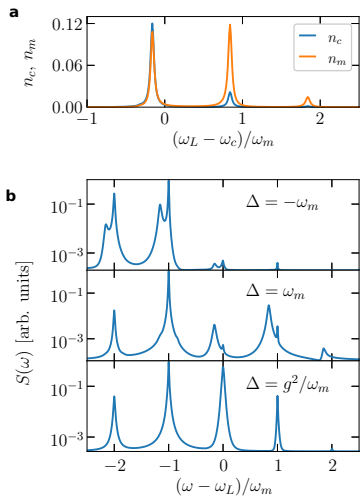}
\caption{
{\bf a} Steady-state photon/phonon populations when the laser frequency is scanned across the cavity resonance. {\bf b} Cavity emitted spectrum for selected detunings of $\Delta{=}\omega_c{-}\omega_L{=}{-}\omega_m,\,\omega_m,\,g^2/\omega_m$ from top to bottom, respectively. The parameters are kept the same as in Fig.~\ref{fig:gscan}(a) (bottom plot) for consistency,
and all simulations are with the more advanced master equation [Eq.~\eqref{eq:rho2}]
at a temperature of $T=4~$K.
\label{fig:detuning} }
\end{figure}

The results above consider the case
where the laser is on resonant with the 
bare cavity resonance, namely
$\omega_L{=}\omega_c$. To have a more
complete picture of detuning dependence,
in Fig.~\ref{fig:detuning} we show the
steady-state populations
as a function of laser detuning,
as well as the emitted spectra
at selected detunings.
All of these simulations are calculated
using the more accurate master equation of Eq.~\eqref{eq:rho2}.  
In Fig.~\ref{fig:detuning}(a), we first plot photon/phonon expectation values ($n_c,\,n_m$) in the steady-state regime.
 Here we define $n_m{=}n_m|_{\rm ss}$
 and $n_c{=}n_c|_{\rm ss}$.
 As can be recognized, the resonance behavior associated with anharmonic energy states are 
 still clearly captured.
 Note, these steady-state cavity populations  can be measured (apart from a constant), e.g., from
 a cavity photoluminescence  experiment, as commonly done for laser excited 
 single quantum dot systems~\cite{Weiler2012}. \
 In Fig.~\ref{fig:detuning}(b), we now consider three different laser detuning cases for the cavity emitted spectrum: $\Delta{=}{-}\omega_m,\,\omega_m,\,g^2/\omega_m$, as labeled on the figure. 
 For $\Delta{=}{-}\omega_m$ ($\omega_L{-}\omega_c{=}\omega_m$), the anharmonic peaks (a-c) are clearly still visible, with a single anti-Stokes peak at $\omega{=}\omega_L{+}\omega_m$. When $\Delta{=}{-}\omega_m$ ($\omega_L{-}\omega_c{=}\omega_m)$, we now easily resolve the d transition as well as the c-transition, and in fact this is the clearest example of resolving the d-transition.
 Finally, $\Delta{=}g^2/\omega_m$ ($\omega_L{-}\omega_c{=}{-}g^2/\omega_m$),
 then we only pick the regular harmonic Raman peaks, though still well into the nonperturbative regime
 (i.e., beyond first order). This later case is not so surprising as we are now
 exciting resonantly with the dressed cavity resonance ($\omega_c{\rightarrow} \omega_c{-}g^2/\omega_m$). Clearly,   a more complete picture is thus obtained by carrying our the emission spectra for a range of detunings, and there is also likely a wide range of coupling scenarios by applying two-color laser fields, i.e., bichromatic driving, and possibly probing heating and cooling effects by varying
 the pump strength and laser detuning.

\section*{Conclusions and Discussion}

\label{sec:con}
 We have introduced and explored the regime of molecular optomechanics in the nonlinear strong coupling regime, where a strong modification of the cavity-emitted spectrum is obtained because of the
influence of higher lying quantum states, which have an anharmonic level spacing $g^2/\omega_m$ as the first excited photon manifold. These nonlinear
anharmonic quantum states can be spectrally resolved if $g^2/\omega_m{>}\kappa$ and $\kappa {<} \omega_m$, which is typically not possible with plasmonic resonators. However,
our cavity design exploited a hybrid metal-dielectric system where a plasmonic dimer is placed on top of a photonic crystal nanobeam cavity. This hybrid design, which is calculated
from first principles, delivers a hybrid mode with a resonance frequency of $\hbar\omega_{c}^{{\rm HQ}}{=}1.61\,{\rm eV}$ and a quality factor of $Q^{{\rm HQ}}{=}3500$ ($\hbar\kappa{=}0.46\,$meV, FWHM). The high $Q$
(small $\kappa$) feature is essential for accessing this regime of optomechanical strong coupling coupling in the sideband-resolved regime. In fact, while the second mode of the same device  has a much higher $g$, the associated
quality factor of $Q^{{\rm LQ}}{=}17$ ($\hbar\kappa=108\,$meV) is too low, which is typical for most plasmonic resonators. Indeed, such broadening fails to reach the sideband resolved regime, despite the fact that the low $Q$ mode has an effective coupling factor of more than two orders of magnitude larger than the high $Q$ mode. However, these low $Q$
pronounced plasmon modes  may be interesting for exploring additional USC and even deep USC effects, which will
be explored in future work.
For our present study,  the sufficiently high $Q$ (low $\kappa$)
and large $g$ are two essential criteria to probe the strong coupling anharmonic ladder states of the optomechical system.
While our designs use extreme small gap antennas, the prospect of using large Raman active and resonant Raman
processes in molecules indicates that emerging experiments 
in quantum plasmonic systems are not too far off reaching such a regime. Different MNP and dielectric-cavity designs could also make the proposals more feasible with larger gap sizes.
We have also shown that the standard master equation generally fails in these regimes, and explored the role
of laser detuning on the steady-state
populations and emission spectra. Indeed our proposed system also allows on to probe dissipation dynamics in the molecular USC regime.

 It is also worth mentioning that recently there have been emerging new designs on dielectric cavity systems with deep sub-wavelength confinement~\cite{SmallV_acs,SmallV_prl}. This could be a major benefit for many cavity-QED applications as very large quality factors are also offered. However, note that for the nonlinear quantum effects studied above, the nonlinear anharmonic energy level shift that is introduced by the strong optomechanical coupling, depends on $\Delta E{=}{-}\hbar g^{2}/\omega_{m}$, and not too critically on the cavity quality factor. Therefore, while these proposed dielectric cavities can offer stronger $Q/V_{c}$ values compared to the particular design introduced here, the mode volume they offer  (and specifically $g^2$) is still significantly smaller than  our hybrid design; having a suitably large $g$ is in fact essential.
 Nevertheless, we anticipate continues improvements
 in both dielectric and plasmonic systems, as well
 as hybrid plasmon-dielectric modes, opening up a wider range of effects in molecular cavity QED and plasmonic cavity QED in general.

\section*{Acknowledgements}
We acknowledge Queen's University and the Natural Sciences and Engineering Research Council (NSERC) of Canada for financial support, and CMC Microsystems for the provision of COMSOL Multiphysics to facilitate this research. We also thank A. Settineri,  S. Savasta, L. Tian and S. Barzanjeh for useful discussions.

\bibliography{Text}

\end{document}